# Giant zero field cooled spontaneous exchange bias effect in phase separated $La_{1.5}Sr_{0.5}CoMnO_6$

J. Krishna Murthy and A.Venimadhav[a]

*Cryogenic Engineering Centre, Indian Institute of Technology, Kharagpur-721302, India.*

**Abstract**

We report a giant zero field cooled exchange bias (ZEB) effect (~0.65 T) in $La_{1.5}Sr_{0.5}CoMnO_6$ sample. Magnetic study has revealed a reentrant spin glass ~90 K, phase separation to spin glass and ferromagnetic phases below 50 K and canted antiferromagnetic transition ~10 K. A small conventional exchange bias (CEB) is established with the advent of spontaneous phase separation down to 10 K. Giant ZEB and enhanced CEB effects are found only below 10 K and are attributed to the large unidirectional anisotropy at the interface of isothermally field induced ferromagnetic phase and canted antiferromagnetic background.

[a] venimadhav@hijli.iitkgp.ernet.in

The exchange bias (EB) is the phenomenon associated with the exchange anisotropy across the ferromagnetic (FM) and antiferromagnetic (AFM) interface[1, 2] and it is ubiquitous to magnetic recording read heads and spintronic devices. In the inhomogeneous magnetic system consisting of hard and soft phases of AFM and FM, such an effect is manifested as a shift in the isothermal M(H) loop with respect to horizontal field axis and vertical magnetization axis.[1,3] Most of the experimental and theoretical studies have revealed that EB effect can be engineered in wide variety of systems by involving inhomogeneous magnetic phases like FM/ferrimagnet, FM/SG and hard/soft FM systems[4-6] in choice of geometries like core-shell nanostructures, granular composites, bilayers and superlattices.[7-10] Lately, there has been a great interest in electrical control of EB devices.[11] In fact C. Vaz *et al*, have taken the advantage of magnetoelectric media to craft the electric field control of the magnetic hysteresis loop.[12]

In contrast to field cooled CEB, in certain systems below the blocking temperature a spontaneous loop shift can be observed without the assistance of external magnetic field and this unusual zero-field-cooled (ZFC) M(H) loop shift is called zero field cooled EB (ZEB) or spontaneous EB effect. Previously, a small exchange bias loop shift after the zero field cooling was considered as an artifact, later Saha *et al*, have provided a model for small but an intrinsic effect in $Ni_{80}Fe_{20}/Ni_{50}Mn_{50}$ system.[13] Recently, a large loop shift after zero field cooling beyond the ambit of experimental artifact has been realized in $NiMnIn$[13] and $Mn_2PtGa$ Heusler alloys and $BiFeO_3$-$Bi_2Fe_4O_9$ nanocomposites.[14-16] It is interesting to note that they all show multiple magnetic phases. The SG phase was found to be important in $NiMnIn$[13] and $BiFeO_3$-$Bi_2Fe_4O_9$ nanocomposite systems for the observation of ZEB effect, however the CEB effect was found to be much larger than the ZEB at a given temperature.[14, 16] While in $Mn_2PtGa$ Heusler compound, the unidirectional anisotropy due to FM clusters embedded in ferrimagnetically ordered matrix seems to be responsible as the possible source of ZEB effect; interestingly, ZEB magnitude was similar to its CEB value.[15] In the case of $Ni_{80}Fe_{20}$/(Ni, Fe)O bilayers, under zero field cooled protocol a positive EB effect was observed and it changes sign after cooling the system under external field.[17] The ZEB effect will be of great interest to electric field control of EB devices as it eliminates the requirement of external magnetic field to create the unidirectional anisotropy. In this letter, we report a giant value of ZEB effect ~0.65 T and a colossal value of CEB effect ~1.17 T in $La_{1.5}Sr_{0.5}CoMnO_6$ (LSCMO) double perovskite. The parent compound $La_2CoMnO_6$ exhibits hard magnetic behavior and multifunctional properties like large magnetoresistance and magnetodielectric effect.[18, 19] Replacing the La by Lu and Y has shown $E^*$-type magnetic ordering and exhibited multiferroic properties.[20, 21] Here we present the observation of FM, SG and phase separation in the hole doped LSCMO and discuss the role of multiple magnetic phases on ZEB and CEB effects.

Polycrystalline LSCMO bulk sample was prepared by conventional sol-gel method. The obtained precursor powder was calcinated at 1300 °C for 24 h. The structural analysis was done by high resolution x-ray diffraction with Cu-K$\alpha$ radiation and its Rietveld refinement using Fullprof suite program has suggested a disordered rhombohedral crystal structure with space group R-3c.[22] Both dc and ac susceptibility magnetic measurements were carried out by using Quantum Design SQUID-VSM magnetometer.

Temperature dependent ZFC ($M_{ZFC}$) and field-cooled ($M_{FC}$) magnetization with ~100 Oe dc field is shown in the Fig. 1(a). Here a paramagnetic (PM) to FM ordering at ~174 K is observed. A magnified view of $M_{ZFC}$ in the inset of Fig. 1(a) shows a decrease in magnetization below a broad magnetic transition ~85 K. Below 50 K an increase in magnetization suggest the increase in FM correlations and this is followed by a canted antiferromagnetic (CAF) transition ($T_{CAF}$) at ~10 K. A large difference between $M_{FC}$ and $M_{ZFC}$ and their bifurcation below ~174 K are the features of glassy behavior. The ac susceptibility (acs) is a superior tool to unveil such magnetic glassy nature, correspondingly, Fig. 1(b) shows the out-of-phase component ($\chi''$) of acs for different frequencies. A sharp and frequency independent peak at ~174 K suggests its FM nature, and a broad frequency dependent peak around ~100 K hints the glassy behavior. However, acs does not show any signature for $T_{CAF}$ ~10 K. The frequency sensitive peak temperature in $\chi''(T)$ fits well to the power law $\tau = \tau_0 (\frac{T_f - T_g}{T_g})^{-zv}$ as shown in inset of Fig. 1(c), where $\tau_0$ is microscopic spin flipping time, $T_f$ is frequency dependent peak temperature, $T_g$ is glassy freezing temperature and zv denotes the critical exponent.[23] The fitted parameters: $\tau_0$ ~4.23x10$^{-11}$ sec, zv ~10.03 ±1.14 and $T_g$ ~90 ±1.36 K indicates that the observed glassy behavior is indeed a reentrant spin glass. In order to understand the collective glassy behavior, we have performed memory effect in acs using standard protocol,[24] Accordingly, first the susceptibility ($\chi'$) was recorded as reference ($\chi'_{ref}$) during the heating mode. In second step, the same measurement was performed with an intermediate waiting time ($t_W$) of ~10$^4$ sec at 75 K, 30 K and 10 K and record the halt magnetic susceptibility ($\chi'_{halt}$) and is shown in the Fig. 1(d). Two dips were observed in the difference curve $\Delta\chi' = \chi'_{ref} - \chi'_{halt}$ as shown in inset of Fig. 1(d). At 75 K, close to SG transition a dip was observed; at 30 K, another broad dip was observed suggesting the coexistence of FM and SG phases below 50 K. However for waiting at 10 K, no dip was observed in $\Delta\chi'$ vs. T curve, this suggests the absence of SG like dynamics below $T_{CAF}$.

System with multiple magnetic phases is potential for obtaining EB effect. The M(H) loop measured in ZFC mode at 2 K is found to be highly asymmetric as shown in the Fig. 2(a). The loop asymmetric along the field axis and magnetization axis can be quantified as EB field ($H_{ZEB}= (|H_{c1}|-|H_{c2}|)/2$) and EB magnetization ($M_{ZEB}=|M_{r1}|-|M_{r2}|)/2$) respectively, $H_{c1}$ and $H_{c1}$ are the positive and negative intercepts of the magnetization curve with the field axis and $M_{r1}$ and $M_{r2}$ are the positive and negative intercepts of the magnetization curve with the magnetization axis respectively. The observed shift in $H_{ZEB}$ and $M_{ZEB}$ is ~0.65 T and ~0.148 $\mu_B$/f.u. respectively. The ZEB reported here is higher than the highest reported so far in $Mn_2PtGa$ Heusler alloys (~0.17 T).[15] In order to explore the effect of field direction on ZEB, we have performed M(H) measurements at 2 K in two protocols[14]: (1) p-type, $0 \rightarrow (+H_{max}) \rightarrow (-H_{max}) \rightarrow 0$ and (2) n-type, $0 \rightarrow (-H_{max}) \rightarrow (+H_{max}) \rightarrow 0$ as shown in the Fig. 2(a). The p-type and n-type M(H) loops measured in ZFC mode are symmetric in nature which indicates that the observed spontaneous shifts in M(H) loops are not experimental artifacts; care has been taken to avoid spurious shift from trapped magnetic field in superconducting magnet by demagnetizing in oscillatory field at 380 K, which is PM regime of LSCMO sample.

The CEB has been measured at 2 K by cooling the system in a magnetic field ($H_{FC}$) of $\pm 5$ T. As shown in the Fig. 2(b), the measured M(H) loop is completely shifted towards the left side for +5 T $H_{FC}$ and it is symmetrically opposite with respect to the measurement performed under –5 T $H_{FC}$, which indicates that the observed EB effect is not originated from the non saturated minor loop. The shift has a giant value of ~1.17 T which is much larger than the CEB of other ZEB systems.[14-16] This value is comparable with the high CEB in ferrimagnetic YbFe$_2$O$_4$ spinel, however, this ferrite does not show ZEB effect.[25]

M(H) measurements at different temperatures is shown in the inset of Fig. 2(c); the squareness of the FM loop diminishes progressively with the decrease of temperature that suggests the dominant AFM interactions at low temperature. Another interesting observation is that the increase in magnetization below 50 K (see inset to Fig. 1(a)) is correlated with the appearance of soft FM phase in the M(H) loops below 50 K (shown by an arrow mark). It can be noticed from Fig. 2(a) that below 10 K, the virgin magnetization curve falls outside the M(H) hysteresis envelop up to a certain field (~3.5 T at 2 K). In order to understand this metamagnetic nature, we have measured isothermal virgin magnetization curves at different temperatures as shown in the Fig. 2(c). Below 10 K, the magnetization increases linearly up to a certain field called critical field ($H_j$) and there after a change in slope with broad transition can be noticed. This behavior is analogous to the field induced metamagnetic transition from CAF to FM phase observed in phase separated systems where the induced FM clusters are kinetically arrested in the AFM matrix.[26]

To get further in sight on field induced growth of FM clusters we have measured magnetization vs. time and is shown in Fig. 2(d). Hear magnetization was measured as a function of time after cooling the sample in zero field to 5 K and soaking it for 100 sec in different dc fields.[26] For low magnetic field (~100 Oe) the magnetization does not show any variation with time. With increasing the magnetic field, the change in magnetization with time also increases and again for a filed higher than $H_j$, it can be noticed that there is no change in magnetization with time. This implicates the growth of field induced of FM clusters up to a certain magnetic field and above which a fraction of CAF phase gets converted to FM phase permanently. This field induced phase separation to FM and AFM phases seems hold the key to understand the EB effect.

In EB systems, repeating the hysteresis loop along with thermal cycling leads to relaxation of uncompensated spin configuration at the interface, consequently M(H) loops become symmetric known as training effect.[2, 27] And smaller this value is better for applications. Since the present sample shows both ZEB and CEB effects, we have investigated the training effect in both the cases at 5 K. For ZEB case, training effect is measured in ZFC mode with several thermal cycles (inset (a) of Fig. 3) and training in CEB is measured after applying 5 T $H_{FC}$ as shown in Fig. 3. Interestingly, we have observed no training of ZEB effect and this indicates that the uncompensated spin configuration at the interface is quite stable against the thermal cycling. On the other hand $H_{CEB}$ decrease with the number of cycles ($\lambda$) and the decrement from $\lambda=1$ to $\lambda=2$ is 4% and to $\lambda=11$ is ~ 10 %; this decrement is very small compared to the giant value of the observed EB effect. This indicates that in LSCMO the metastable spin configuration at the interface is reasonably stable even against the applied magnetic field like in CEB systems.[3] The decrement in CEB is normally attributed to the switching and orientation of pinned AFM spins at the interface after the magnetization reversal of FM layer during consecutively cycles.[28] The decrease in $H_{CEB}$ or $M_{EB}$ can be fit to the following empirical power law for $\lambda \geq 2$,

$$H_{CEB}(\lambda) - H_{CEB}^{\infty} \alpha \frac{1}{\sqrt{\lambda}} \qquad \text{-------- (1)}$$

where, $H_{CEB}^{\infty}$ is the value of $H_{CEB}$ for $\lambda=\infty$. The solid line indicates the satisfactory fit to Eqn. (1), for the decreasing of $H_{CEB}$ with number of cycles and obtained fitting parameter, $H_{CEB}^{\infty}$ = 7.46 kOe. Binek has proposed a recursive formula in the framework of spin configurational relaxation to comprehend the training effect which describes the $\lambda+1^{th}$ loop shift with the $\lambda^{th}$ one as $H_{CEB}(\lambda+1)-H_{CEB}(\lambda) = -\gamma[H_{CEB}(\lambda)-H_{CEB}(\lambda=\infty)]$ [3,27] and the obtained sample dependent constant, $\gamma$ is ~3.87 X $10^{-7}$.

The temperature dependence of coercive field ($H_c$) and remnant magnetization ($M_r$) are shown in the Fig. 4(a). Here, below $T_C$, both $H_c$ and $M_r$ increases with the decrease of temperature and both show a sudden down turn below the CAF transition. The temperature dependence of both ZEB and CEB effects are shown in the Fig. 4(b) and it can be noticed that ZEB effect exists only below 10 K. Interestingly, below 10 K, the magnetic ground state prefers CAF state and the application of magnetic field above $H_j$ leads to phase separation. In CEB, the temperature evolution of $H_{CEB}$ (with 5 T of $H_{FC}$) starts below 50 K with the advent of spontaneous phase separation to SG and FM phases. Below 10 K, CEB follows the ZEB effect and a sharp increase by an order of magnitude can be noticed. The $H_{FC}$ dependence of $H_{CEB}$ and $M_{CEB}$ in the FC mode at 5 K is shown in the inset of Fig. 4(b). With increasing $H_{FC}$, $H_{CEB}$ increases steeply up to 0.1 T and shows a small increment until 3 T and above that no noticeable change is observed. Similarly $M_{CEB}$ also increases up to 3 T and remains unchanged thereafter. The variation in $H_{CEB}$ and $M_{CEB}$ at field ~3 T evidently coincides with the critical field $H_j$.

The CEB is widely studied and the extent of this effect depends on the architecture of the FM/AFM interface. Recently, a small CEB (~130 Oe) effect was reported in Sr doped isostructural $La_2NiMnO_6$ system and was ascribed to the exchange interaction between FM phase and AFM antiphase boundaries.[29] In the present LSCMO, upon hole doping, various AFM interactions like $Co^{3+}$-O-$Co^{2+}$, $Co^{2+}$-O-$Co^{2+}$ and $Co^{3+}$-O-$Co^{3+}$ become dominant with the decrease of thermal energy and consequently breaks the long range FM ordering of $Co^{2+}$-O-$Mn^{4+}$ that leads to reentrant spin glass, spontaneous phase separation and field induced phase separation at low temperatures. The observed CEB effect between 10 K-50 K is due to the unidirectional anisotropy at the FM and SG interface in a way similar to the spontaneously phase separated cobaltites and manganites.[3] However, the unusual giant ZEB and CEB effects below the CAF transition are exceptional to LSCMO system.

In $NiMnIn_{13}$ Heusler alloys and $BiFeO_3$-$Bi_2Fe_4O_9$ nanocomposites, it was made clear that the SG phase plays the central role in obtaining ZEB effect.[14, 16] Contrastingly, in LSCMO, the SG phase transition is at much higher temperature than the ZEB effect and the giant value of the effect hints at uncompensated interface spin structure. A qualitative understanding can be drawn from the Fig. 4(c), where the isothermally field induced variations in spin configuration below CAF transition is shown. With ZFC, system undergoes to CAF ordering for T < 10 K. During the initial magnetization, for magnetic field strength higher than $H_j$ induces a soft FM phase (blue shaded regions). While decreasing the field (in the second cycle) the field induced FM phase is kinetically arrested and coexists with CAF matrix that creates a large unidirectional anisotropy at their interface (shown as red lines) in the Fig. 4(c). This field induced phase separation is significant and spin reversal becomes expensive due to the newly formed interfaces. The M(H) loop can get shifted to either negative or positive field axis depends on the initial direction of the applied field (i.e., p-type and n-type M(H) loops). In the field cooled case or CEB, there exist FM clusters even at H= 0 below the CAF transition. And further isothermal field ramping produces more such clusters that increases the number of FM/AFM interfaces which enhances the EB field. A close look reveals that below CAF transition, the CEB is actually an enhanced effect of the ZEB.

In summary, various magnetic measurements have revealed multiple magnetic phases and both CEB and ZEB effects in LSCMO system. The CEB effect in the temperature range of 10-50 K has been ascribed to the interfacial exchange anisotropy between the spontaneously phase separated FM and SG phases. Metamagnetic behaviour and a field induced phase separation below CAF transition are responsible for the observed giant ZEB and CEB effects. The observed ZEB would greatly benefit the spin reversal devices if the loop shift in the unmagnetized state can be electrically controlled.

**Acknowledgements**

We sincerely thank M.G. Blamire for the thoughtful discussions. The authors also acknowledge IIT Kharagpur for funding VSM-SQUID magnetometer. Krishnamurthy thanks CSIR-UGC, Delhi for SRF.

**References:**
1. W. H. Meiklejohn and C. P. Bean, Phys. Rev. 102, 1413 (1956).
2. J. Nogués and I. K. Schuller, J. Magn. Magn. Mater. 192, 203 (1999).
3. S. Giri, M. Patra and S. Majumdar, J. Phys.: Condens. Matter 23, 073201 (2011).
4. Q. K. Ong, A. Wei, and X.-M. Lin, Phys. Rev. B 80, 134418 (2009).
5. M. Ali, P. Adie, C. H. Marrows, D. Greig, B. J. Hickey, and R. L. Stamps, Nature Mater. 6, 70 (2007).
6. D. Navas, J. Torrejon, F. Béron, C. Redondo, F. Batallan, B. P. Toperverg, A. Devishvili, B. Sierra, F. Castaño, K. R. Pirota, and C. A. Ross, New J. Phys. 14, 113001 (2012).

7. J. A. De Toro, J. P. Andrés, J. A. González, P. Muñiz, T. Muñoz, P. S. Normile, and J. M. Riveiro, Phys. Rev. B 73, 094449 (2006).
8. Z. M. Tian, S. L. Yuan, S. Y. Yin, L. Liu, J. H. He, H. N. Duan, P. Li, and C. H. Wang, Appl. Phys. Lett. 93, 222505 (2008).
9. S. K. Mishra, F. Radu, H. A. Dürr, and W. Eberhardt, Phys. Rev. Lett. 102, 177208 (2009).
10. M. Gibert, P. Zubko, R. Scherwitzl, J. Íñiguez, and J. M. Triscone, Nature Mater. 11, 195 (2012).
11. X. He, Y. Wang, N. Wu, A.N. Caruso, E. Vescovo, K. D. Belashchenko, P.A. Dowben, and C. Binek, Nature Mater. 9, 579 (2010).
12. C. A. F. Vaz, J. Phys.: Condens.Matter 24, 333201 (2012).
13. J. Saha and R. H. Victora, Phys. Rev. B 76, 100405(R) (2007).
14. B. M. Wang, Y. Liu, P. Ren, B. Xia, K. B. Ruan, J. B. Yi, J. Ding, X. G. Li, and L. Wang, Phys. Rev. Lett. 106, 077203 (2011).
15. A. K. Nayak, M. Nicklas, S. Chadov, C. Shekhar, Y. Skourski, J. Winterlik, and C. Felser, Phys. Rev. Lett. 110, 127204 (2013).
16. T. Maity, S. Goswami, D. Bhattacharya, and S. Roy, Phys. Rev. Lett. 110, 107201 (2013).
17. H. Ouyang, K.-W. Lin, C. –C. Liu. S.-C. Lo, Y. –M. Tzeng, Z. –Y. Guo, and J. van Lierop, Phys. Rev. Lett. 98, 097204 (2007).
18. M. Viswanathan, P. S. Anil Kumar, V. S. Bhadram, C. Narayana, A. K. Bera and S. M. yusuf, J. Phys.: Condens. Matter 22, 346006 (2010).
19. J. K. Murthy and A. Venimadhav, J. Appl. Phys. 111, 024102 (2012).
20. S. Y. Vilar and C. D. Batista et al, Phys. Rev. B 84, 134427 (2011).
21. G. Sharma, J. Saha, S. D. Kaushik, V. Siruguri, and S. Patnaik, Appl. Phys. Lett. 103, 012903 (2013).
22. I. O. Troyanchuk, A. P. Sazonov, H. Szymczak, D. M. Többens, and H. Gamari-Seale, J. Exp. and Theor. Phys. 99, 363 (2004).
23. J. A. Mydosh, *Spin Glasses: An Experimental Introduction* (Taylor & Francis, London, 1993).
24. K. Jonason, E. Vincent, J. Hammann, J. P. Bouchaud, and P. Nordblad, Phys. Rev. Lett, 81, 3243 (1998).
25. Y. Sun, J. -Z. Cong, Y. -S. Chai, L. -Q. Yan, Y. -L. Zhao, S. -G. Wang, W. Ning, and Y. -H. Zhang, Appl. Phys. Lett. 102, 172406 (2013).
26. D. S. Rana, D. G. Kuberkar, and S. K. Malik, Phys. Rev. B 73, 064407 (2006).
27. C. Binek, Phys. Rev. B 70, 014421 (2004).
28. X. P. Qiu, D. Z. Yang, S. M. Zhou, R. Chantrell, K. O' Grady, U. Nowak, J. Du,X. J. Bai and L. Sum, Phys. Rev. Lett. 101, 147207 (2008).
29. Y, Guo, L. Shi, S. Zhou, J. Zhao, C. Wang, W. Liu and S. Wei, J. Phys. D: Appl. Phys. 46, 175302 (2013).

**Figure captions:**

**FIG. 1: a)** M(T) under ZFC and FC protocols, inset shows the magnified view of ZFC magnetization at low temperatures, **(b)** temperature variation of $\chi''(T)$ for different frequencies **(c)** the power law fit of $\tau$ vs. $T_f$ and **(d)** temperature variation of $\chi'_{ref}$ and $\chi'_{halt}$; inset shows the $\Delta\chi'$ vs. T(K) plot,

**FIG. 2: (a)** M(H) loops measured at 2 K in ZFC mode, dotted lines are the virgin magnetization curves, **(b)** M(H) loops at 2 K with $H_{FC} = \pm 5$ T, **(c)** the 1$^{st}$ and 2$^{nd}$ branch of M(H) loops at different temperatures and the inset shows the magnified view of isothermal M(H) measurements at different temperatures in ZFC mode, and **(d)** time dependence of magnetization for different dc fields at 5 K.

**FIG. 3:** Training effect of CEB at 5 K, inset shows **(a)** training effect of ZEB at 5 K and **(b)** $H_{CEB}$ vs. $\lambda$, here solid line show the fitting of the experimental data to Eqn. (1).

**FIG. 4:** Temperature dependence of **(a)** $H_c$ and $M_r$; **(b)** $H_{ZEB}$ and $H_{CEB}$ and the inset show the $H_{FC}$ dependence of conventional $H_{CEB}$ and $M_{CEB}$ at 5 K, and (c) schematic diagram of isothermal magnetization process with spin configurations for zero field cooled (ZEB) and field cooled (CEB) cases at $T < T_{CAF}$.

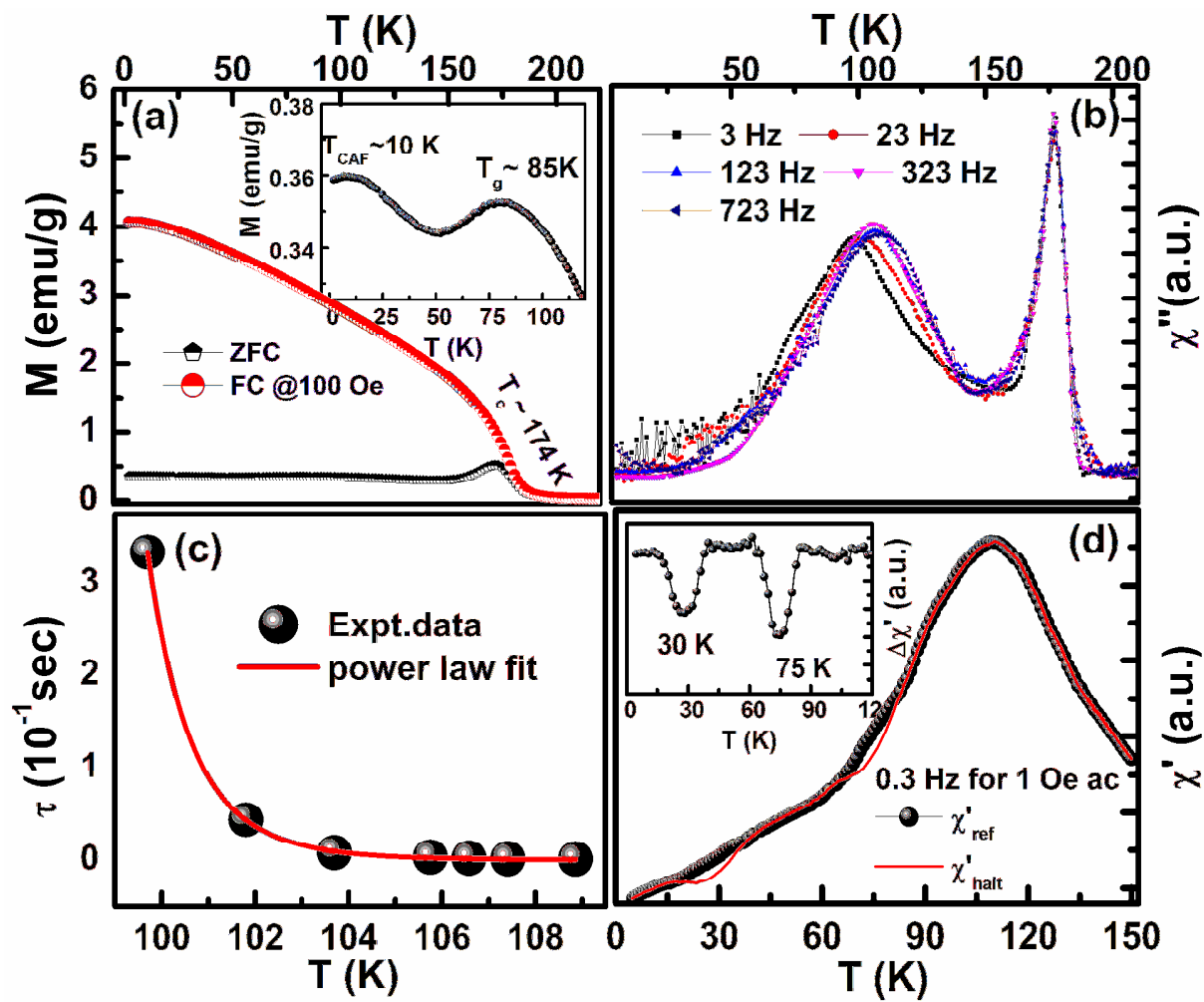

Fig 1

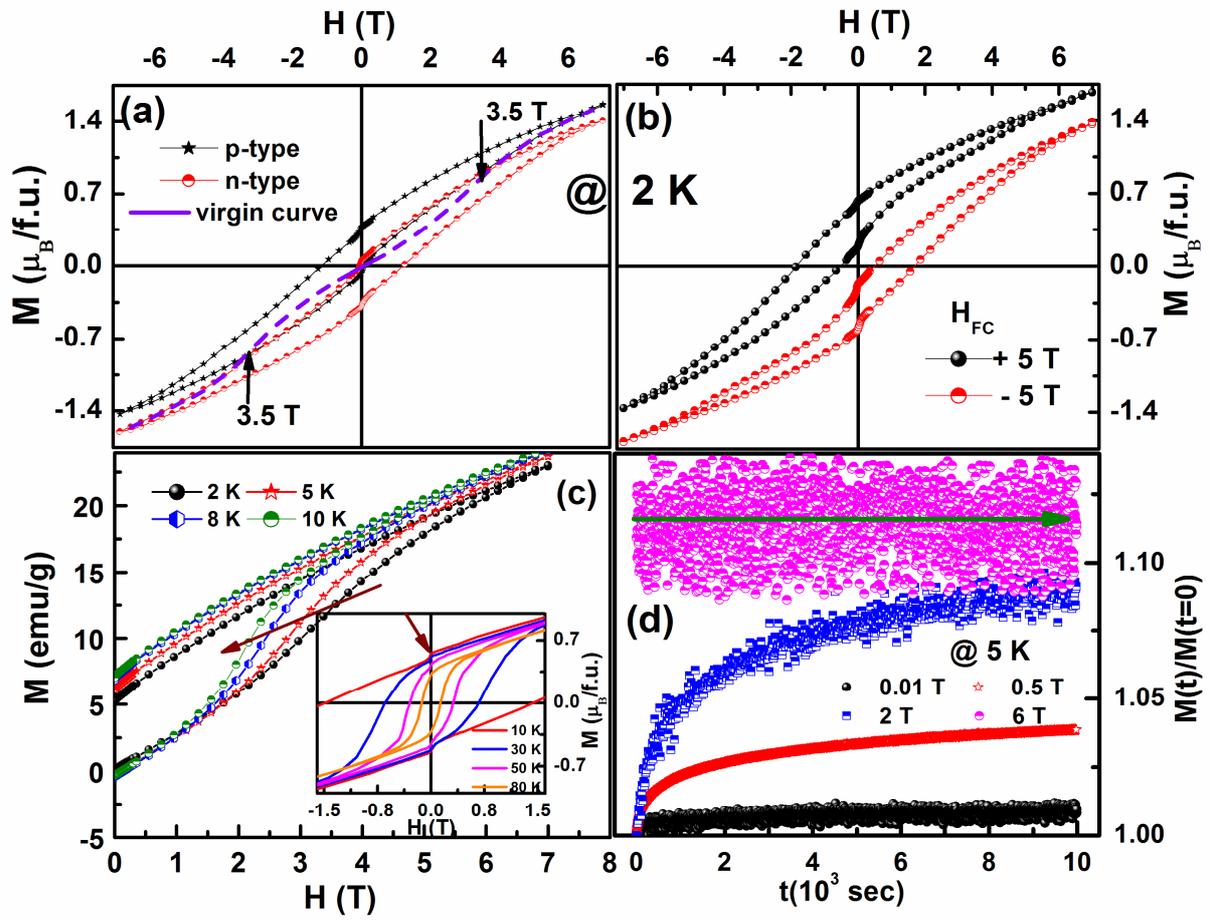

Fig 2

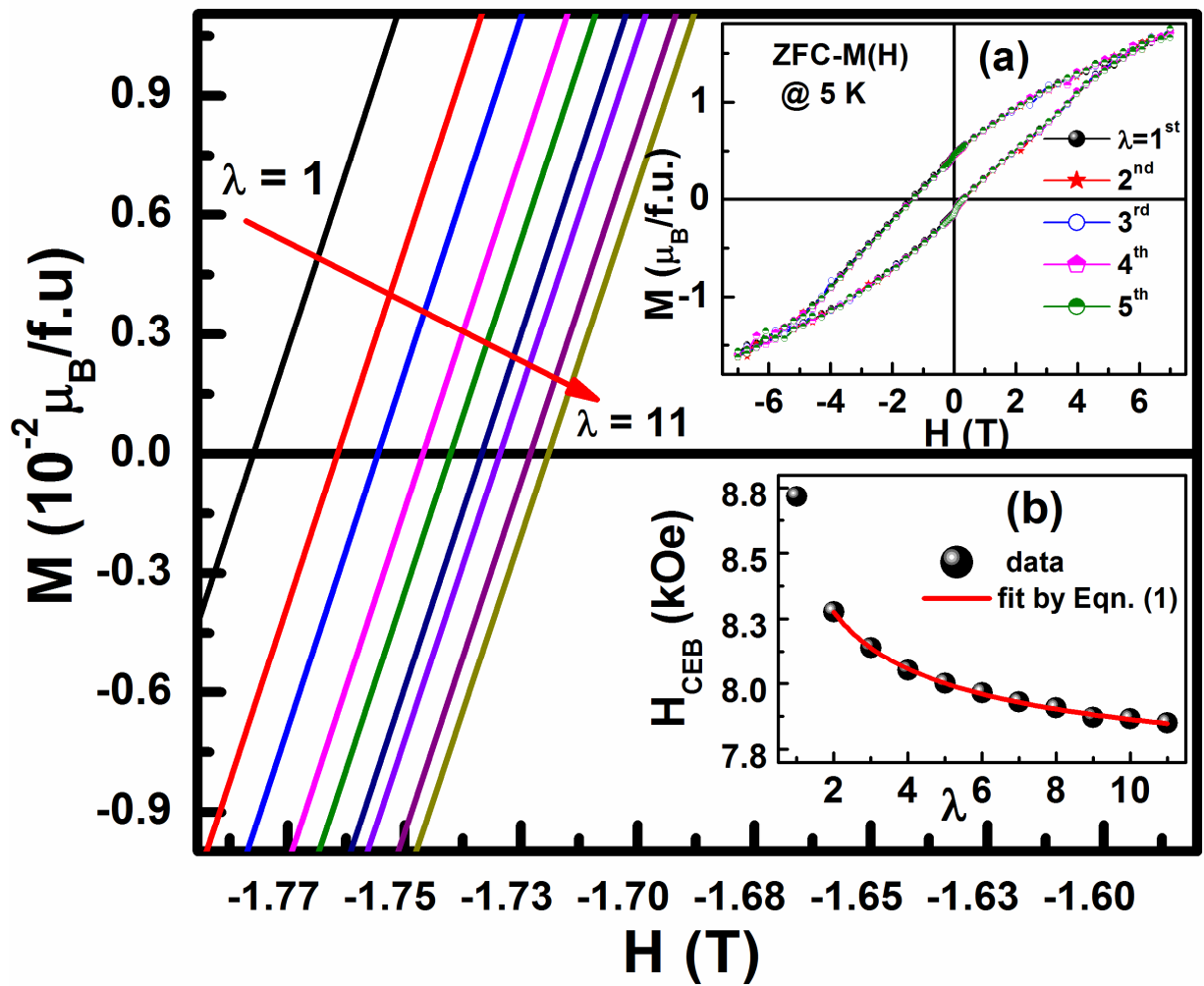

Fig 3

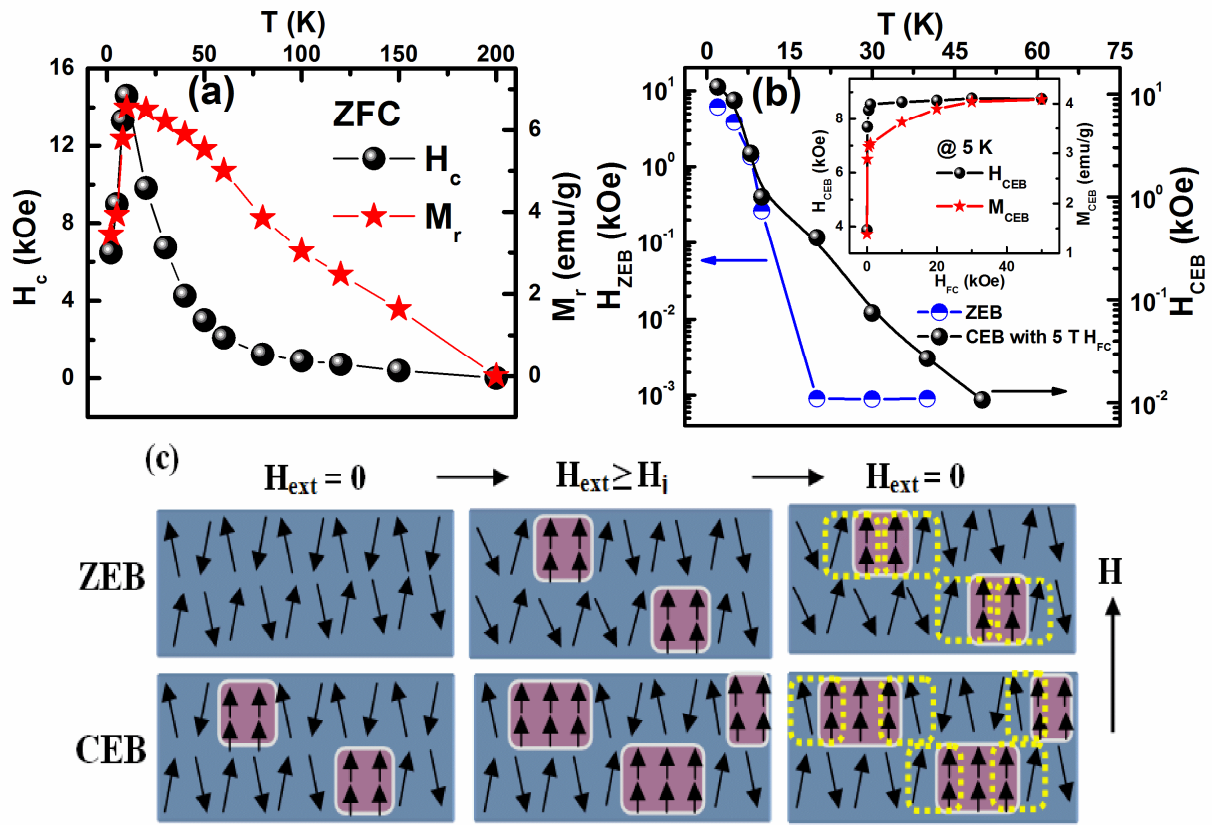

Fig 4